\documentclass[fleqn,usenatbib]{mnras}

\usepackage{newtxtext,newtxmath}

\usepackage[T1]{fontenc}

\DeclareRobustCommand{\VAN}[3]{#2}
\let\VANthebibliography\thebibliography
\def\thebibliography{\DeclareRobustCommand{\VAN}[3]{##3}\VANthebibliography}

\usepackage{graphicx}	
\usepackage{amsmath}	

\usepackage{dcolumn}
\usepackage{ulem}
\usepackage{bm}

\newcommand{\mrm}[1]{\mathrm{#1}}
\newcommand{\nuc}[2]{$\mrm{^{#2}#1}$}

\usepackage{soul}
\usepackage{xcolor}

\title[Sub-GeV Dark Matter Annihilation]{Sub-GeV Dark Matter Annihilation:\\Limits from Milky Way observations with INTEGRAL}

\author[Siegert, Calore \& Serpico]{
Thomas Siegert,$^{1}$\thanks{E-mail: thomas.siegert@uni-wuerzburg.de}
Francesca Calore,$^{2}$
and Pasquale Dario Serpico$^{2}$
\\
$^{1}$Julius-Maximilians-Universität Würzburg, Fakultät für Physik und Astronomie, Institut für Theoretische Physik und Astrophysik,\\~Lehrstuhl für Astronomie, Emil-Fischer-Str. 31, D-97074 Würzburg, Germany\\
$^{2}$LAPTh, CNRS, USMB, F-74940 Annecy, France
}

\date{Accepted XXX. Received YYY; in original form ZZZ}

\pubyear{2023}

\begin{document}
\label{firstpage}
\pagerange{\pageref{firstpage}--\pageref{lastpage}}
\maketitle

\begin{abstract}
From 16 years of INTEGRAL/SPI $\gamma$-ray observations, we derive bounds on annihilating light dark matter particles in the halo of the Milky Way up to masses of about 300\,MeV.
We test four different spatial templates for the dark matter halo, including a Navarro-Frenk-White (NFW), Einasto, Burkert, and isothermal sphere profile, as well as three different models for the underlying diffuse Inverse Compton emission.
We find that the bounds on the s-wave velocity-averaged annihilation cross sections for both the electron-positron and the photon-photon final states are the strongest to date from $\gamma$-ray observations alone in the mass range $\lesssim 6$\,MeV.
We provide fitting formulae for the upper limits and discuss their dependences on the halo profile.
The bounds on the two-photon final state are superseding the limits from the Cosmic Microwave Background in the range of 50\,keV up to $\sim 3$\,MeV, showing the great potential future MeV mission will have in probing light dark matter.
\end{abstract}



\section{Introduction}\label{sec:level1}
The dark matter (DM) phenomenon emerges from many different astrophysical and cosmological observations.
Despite decades of direct and indirect searches for particle DM, its true nature appears more elusive than expected from the `WIMP miracle' \citep{Jungman1996_susydm}, which finds support in theories predicting weak scale DM candidates, such as neutralinos in supersymmetric extensions of the Standard Model for particle physics.
If they had been produced thermally in a hot Big Bang and in order to match the present DM abundance, we know that their cross sections with the Standard Model sector should match the so-called \textit{thermal relic} value \citep[e.g.,][]{Steigman2012_thermalrelic}. 

Nonetheless, there is no fundamental reason as to why DM particles \citep[if they are indeed particles; e.g.,][]{Siegert2022_RetII,Berteaud2022_SPI_PBH} should be found in the electroweak mass range.
The past decade witnessed a broadening of the theory landscape for particle DM, which has gone along with impressive technological developments allowing to enlarge the exploitable parameter space.
In particular, models introducing light (i.e., with masses below 1\,GeV) DM candidates have been put forward in the context of portals' models where DM and Standard Model particles interactions are mediated by messenger particles, such as new vector or scalar particles \citep[see][for an updated overview on the theory and experimental aspects of light DM]{2023arXiv230501715A}.
Also in this case thermal production is possible, from eV up to GeV masses, although severely constrained below 1\,MeV by Big Bang Nucleosynthesis (BBN) and Cosmic Microwave Background (CMB) observations \citep{Slatyer2016_CMB_DM,2020JCAP...01..004S,2021JCAP...08A..01S}.
Complementary constraints came from $\gamma$-ray observations of the diffuse MeV emission \citep[e.g.,][]{Essig2013_DMlimits_gammarays, Laha2020_PMBHDM,2021PhRvD.103f3022C,2023JCAP...07..026C}, for both self-annihilating and decaying DM candidates.
It is important to obtain multiple constraints on these light particles, since different observables are sensitive to DM at different epochs, with different velocity distributions in case of non-trivial velocity dependences, and rely on different assumptions.
For instance, to the best of our knowledge, BBN bounds have not been derived when both new light DM and a low reheating temperature are simultaneously assumed.
If observables are sensitive to very low cross sections, one could additionally probe non-thermal light DM production scenarios.

In this paper, we focus on the possible self-annihilation of DM particles in the mass range between 50\,keV and 300\,MeV.
It is the third study in a sequence that uses a 16-year data set from the $\gamma$-ray spectrometer telescope SPI \citep{Vedrenne2003_SPI} onboard the INTEGRAL satellite \citep{Winkler2003_INTEGRAL}.
Recently in \citet{Siegert2022_MWdiffuse}, we measured the diffuse emission spectrum in the Milky Way from 0.5--8\,MeV, which was used by \citet{Berteaud2022_SPI_PBH} to study the possible evaporation of primordial black holes as DM candidates, and by \citet{Calore2023_lightDM} to set bounds on decaying light DM particles, such as axion-like particles and sterile neutrinos, leading to the strongest constraints to date on these candidates.
In particular, with our previous works, we demonstrated that the spatial information of the DM signal, i.e. the chosen template(s), is key in setting robust and unbiased constraints on DM, besides being the only way to guarantee not to miss a possible signal in the data.

For annihilating DM particles, the emitted $\gamma$-ray signal depends on the integral along the line of sight of the DM density squared (two-particle process), which implies that the DM emission template is very different from the one of decaying particles \citep[see, for example, our use case in][]{Calore2023_lightDM}.
Because the squared DM density profiles are much more peaked compared to the non-squared profiles, the $\gamma$-ray data analysis becomes more complex:
The angular resolution of INTEGRAL/SPI of $2.7^\circ$ with extended wings out to $\approx 10^\circ$ \citep{Attie2003_SPI} will mimic point-like behaviour for such halo profiles even though it spans the entire region of interest.
Thus, there will be stronger correlations of real point sources with a squared halo profile, which requires a new, adapted, analysis of the data compared to non-squared density profiles \citep{Berteaud2022_SPI_PBH,Calore2023_lightDM}.
Still, INTEGRAL/SPI is currently the only soft $\gamma$-ray instrument which can be used to estimate and to set limits on properties for particle DM annihilations in the MeV--GeV range in addition to CMB measurements \citep{Slatyer2016_CMB_DM}.

This paper is structured as follows:
In Sec.\,\ref{sec:data}, we recapitulate on the INTEGRAL/SPI dataset used in this and previous works.
Sec.\,\ref{sec:models} describes the DM density halo models and the spectral functions for the data analysis.
Our results, separated into final states of two photons and electron-positron pairs, as well as for different spectral modelling assumptions, are presented in Sec.\,\ref{sec:results}.
We discuss our findings in the context of existing studies and DM models in Sec.\,\ref{sec:discussion}.

\section{Gamma-ray dataset}\label{sec:data}
We use the INTEGRAL/SPI data set from \citet{Siegert2022_MWdiffuse} to describe the soft $\gamma$-ray emission in the energy band between 0.05 and 8\,MeV.
The data set includes about 36000 pointed observations of 1900\,s each from 16 years between 2003 and 2019 in the selected region of interest, covering longitudes $|\ell| \leq 47.5^\circ$ and latitudes $|b| \leq 47.5^\circ$.
Due to the strong degeneracy between individual spatial components below 50\,keV photon energies, we use 22 logarithmically spaced energy bins from 51 to 8000\,keV, interrupted by introducing the strong $\gamma$-ray lines at 511\,keV (6\,keV broad) and 1809\,keV (8\,keV).
We calculate our spatial models on a cartesian pixel grid of $0.5^\circ \times 0.5^\circ$, totalling 36100 pixels.
In this way, we oversample the instrumental resolution of SPI of $2.7^\circ$, which helps to distinguish diffuse but peaked emission from point-like sources.
The instrumental background is modelled self-consistently by the methods developed and established in previous works \citep{Diehl2018_BGRDB,Siegert2019_SPIBG}.
For more details about the dataset we refer the reader to \citet{Siegert2022_MWdiffuse}.

We also remind the reader that we adopt a two-step analysis:
First, a spatial decomposition based on template models is performed, so to extract spectral data points, which are then used in a second stage to run a spectral fit.
In both steps, the DM annihilation signal is included.

\section{Astrophysical models}\label{sec:models}
\subsection{Spatial templates}
\subsubsection{Galactic fore- and back-ground emission\label{sec:astro_foreground}}
Given our dataset, ranging from 0.05–8\,MeV, the Galactic back- and fore-ground emission has to be taken into account to potentially extract the spectrum of a DM halo template in this energy band.
Similar to our previous works, we take into account the following components:
A set of resolved sources from 50\,keV to $\approx 1$\,MeV whose total number decreases as a function of energy because their individual flux levels become weaker, modelled as point sources at their known position, assuming their fluxes are constant in time.
The unresolved point sources, mainly cataclysmic variables, cannot be distinguished separately but contribute in the range up to $\lesssim 100$\,keV \citep{Bouchet2008_imaging}.
Positron annihilation, of both ortho- ($\leq 511$\,keV) and para-Positronium (around 511\,keV), is modelled by the four-component template of \citet{Siegert2016_511}.
The \nuc{Al}{26} decay $\gamma$-ray line, at 1809\,keV, is modelled only in a single energy bin from 1805--1813\,keV with the SPI \nuc{Al}{26}-map from \citet{Bouchet2015_26Al}.
Finally, we model the Inverse Compton (IC) emission in the Milky Way in the entire band from 0.05--8\,MeV with three assumptions, briefly explained in the following, to estimate systematic uncertainties \citep[see also][for more systematic uncertainty estimates]{Siegert2022_MWdiffuse}:

The baseline IC model used in this work is adopted from \citet{Bisschoff2019_Voyager1CR}, taking into account the Voyager 1 measurements \citep{Stone2013_Voyager1_CR}.
We use GALPROP v56 \citep[e.g.,][]{Strong2011_GALPROP} to calculate the IC emission in our band using the electron spectrum derived in \citet{Bisschoff2019_Voyager1CR}.
In addition, we calculate two more variants of the IC spectrum, 1) unifying the spectral index of the electron rigidity spectrum to $\delta_1 = \delta_2 = \delta = 0.5$ as this model fits the SPI data above 500\,keV best \citep{Siegert2022_MWdiffuse}, and 2) enhancing the amplitude of the optical component of the interstellar radiation field by a factor of 10 as it was shown previously to better match MeV observations \citep{Bouchet2011_diffuseCR}.
For details about the Galactic fore- and back-ground components, we refer again to our previous studies \citep{Siegert2022_MWdiffuse,Berteaud2022_SPI_PBH,Calore2023_lightDM}.

\subsubsection{Dark matter halo profiles}\label{sec:halos}
We take into account the variety of DM halo profiles discussed in the literature, benchmarking possible radial behaviours, and perform our analysis (Sec.\,\ref{sec:results}) for four different DM halo assumptions, combined with three assumptions on the underlying IC emission as explained above.
For reference, we use a distance of the Solar System to the Galactic centre of $d_\odot = 8.178$\,kpc, 
and a DM density at the solar circle of $13.17 \times 10^{-3}\,\mrm{M_{\odot}\,pc^{-3}} \equiv 0.5\,\mrm{GeV\,cm^{-3}}$ \citep{Benito2021_DMhalo}.
The conversion factor between $\mrm{M_{\odot}\,pc^{-3}}$ and $\mrm{GeV\,cm^{-3}}$ is 38 in natural units.

\begin{table}
	\centering
	\begin{tabular}{c|ccccc|c}
		Profile & $\rho_0$ & $r_s$ & $\alpha$ & $\beta$ & $\gamma$ & $\log_{10}(J)$  \\
		\hline
		NFW & $10.69$ & $20$ & $1$ & $3$ & $1$ & $22.85$ \\
		EIN & $2.506$ & $20$ & $0.17$ & $-$ & $-$ & $23.03$ \\
        BUR & $21.66$ & $20$ & $-$ & $-$ & $-$ & $22.30$ \\
		ISO & $17.94$ & $20$ & $2$ & $4$ & $0$ & $22.26$ \\
		\hline
		\hline
		
	\end{tabular}
	\caption{Definition of DM density profiles and $J$-factors in the region of interest $|\ell|\leq47.5^{\circ}$, $|b|\leq47.5^{\circ}$. The profiles are calculated according to Eqs.\,(\ref{eq:NFW_profile})--(\ref{eq:einasto_profile}), and line-of-sight-integrated via Eq.\,(\ref{eq:los_integration}). The units are, from left to right, $10^{-3}\,\mrm{M_\odot\,pc^{-3}}$, $\mrm{kpc}$, $1$, $1$, $1$, and $\log_{10}\left[\mrm{GeV^2\,cm^{-5}}\right]$, respectively. Note that the parameter $\alpha$ has a different meaning for the Einasto profile.}
	\label{tab:DM_profiles}
\end{table}

\begin{figure}
    \centering
    \includegraphics[width=\columnwidth]{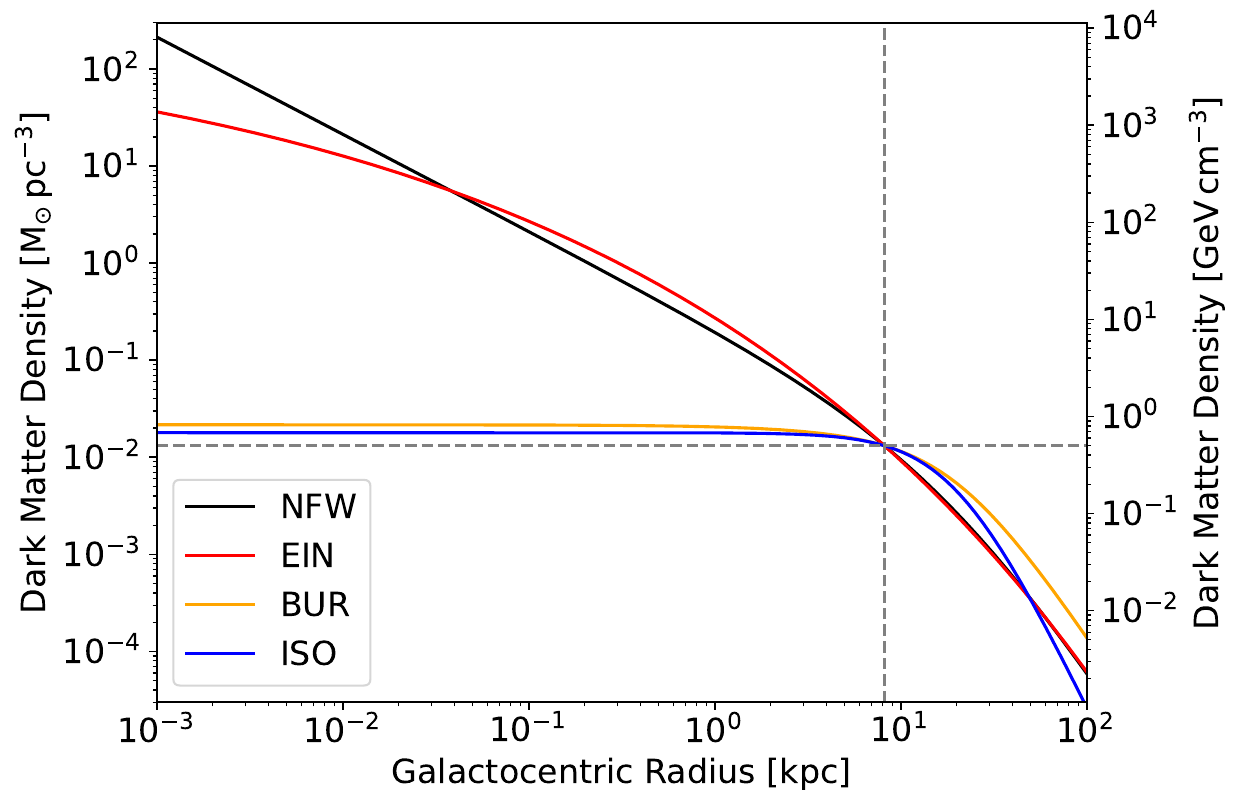}
    \caption{Radial profiles of the DM density models used in this work. The values at the Solar circle are marked with dashed lines.}
    \label{fig:density_profiles}
\end{figure}

We use the general double power-law profile, or Zhao profile \citep{1997MNRAS.287..525Z}, Eq.\,(\ref{eq:DP_profile}), with five parameters, namely the specific DM density, $\rho_0$, the scale radius, $r_s$, and three shape parameters, $\alpha$, $\beta$, and $\gamma$, to construct either a Navarro-Frenk-White profile \citep[$\alpha = 1$, $\beta = 3$, $\gamma = 1$; NFW;][]{Navarro1997_NFW}, or an isothermal sphere profile ($\alpha = 2$, $\beta = 4$, $\gamma = 0$; ISO).
\begin{eqnarray}
    \rho_{\rm DP}(r;\rho_0,r_s,\alpha,\beta,\gamma) & = & \frac{\rho_0}{\left(r/r_s\right)^{\gamma}\left[1 + \left(r/r_s\right)^{\alpha} \right]^{(\beta - \gamma)/\alpha}} \label{eq:DP_profile}\\
    \rho_{\rm NFW}(r;\rho_0,r_s) & = &  \frac{\rho_0}{\left(r/r_s\right)\left[1 + \left(r/r_s\right) \right]^{2}} \label{eq:NFW_profile}\\
    \rho_{\rm ISO}(r;\rho_0,r_s) & = &  \frac{\rho_0}{\left[1 + \left(r/r_s\right)^{2} \right]^{2}} \label{eq:ISO_profile}
\end{eqnarray}
In addition, we use the empirical Burkert profile \citep[BUR;][]{Burkert1995_dm},
\begin{equation}\label{eq:burkert_profile}
	\rho_{\rm BUR}(r;\rho_0,r_s) = \frac{\rho_0 r_s^3}{\left(r_s + r\right)\left(r_s^2 + r^2\right)}\mrm{,}
\end{equation}
which is less steep at larger radii, and the Einasto profile \citep[EIN;][]{Einasto1965_Einasto},
\begin{equation}\label{eq:einasto_profile}
	\rho_{\rm EIN}(r;\rho_0,r_s,\alpha) = \rho_0 \exp\left( - \frac{2}{\alpha} \left[\left(\frac{r}{r_s}\right)^{\alpha} - 1 \right] \right)\mrm{,}
\end{equation}
which avoids the cusp in the centre of the Galaxy.
The exact parameters used for the halo profiles are in accordance with uncertainties of literature values~\citep{Benito2021_DMhalo}, and listed in Tab.~\ref{tab:DM_profiles}.
We note that the parameter $\alpha$ has a different meaning for the EIN profile compared to the other DM density profiles.
We show the spherically symmetric profiles in 1D in Fig.~\ref{fig:density_profiles}.

\begin{figure*}
    \centering
    \includegraphics[width=2\columnwidth]{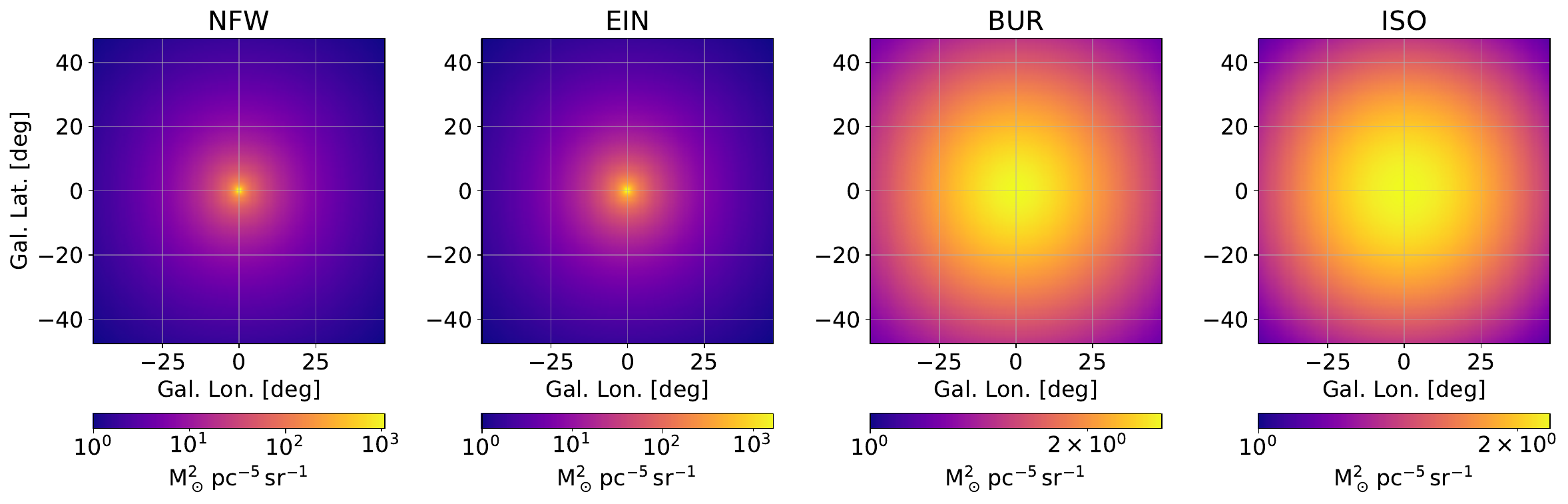}
    \caption{Line-of-sight-integrated DM density halo profiles, squared, in absolute units of $\mrm{M_{\odot}^2\,pc^{-5}\,sr^{-1}}$. The pixel sizes are $0.5^\circ \times 0.5^\circ$, with a total region of interest of $\Delta \Omega = 2.44\,\mrm{sr}$.}
    \label{fig:2D_profiles}
\end{figure*}

The line-of-sight integration is performed according to 
\begin{equation}
    J = \frac{1}{\Delta \Omega} \int_{\Delta\Omega}\int_0^{+\infty}\,\rho^2(s)\,ds\,d\Omega\mrm{,}
    \label{eq:los_integration}
\end{equation}
for DM annihilation, where the units of $J$ are $\mrm{GeV^2\,cm^{-5}}$ or $\mrm{M_{\odot}^2\,pc^{-5}}$.
The 2D image templates in Fig.\,\ref{fig:2D_profiles} show the differential $J$-factors as a function of Galactic coordinates, that is, per steradian of pixel size $0.5^\circ \times 0.5^\circ \approx 7.6 \times 10^{-5}\,\mrm{sr}$ at latitudes $b = 0^\circ$, up to $\approx 5.2 \times 10^{-5}\,\mrm{sr}$ at latitudes $|b| = 47.5^\circ$.
We anticipate that the DM $\gamma$-ray flux is directly proportional to the geometrical, $J$-factor, as per Eq.\,(\ref{eq:astro_spec}).

\subsection{Spectral models}\label{sec:spectral models}
\subsubsection{Galatic fore- and back-ground emission}
The spectral model used for fitting the total Galactic spectrum from 0.05--8\,MeV is provided in Eq.\,(\ref{eq:total_spectral_model}),
\begin{eqnarray}
    \left(\frac{dN}{dE\,dA\,dt}\right)_{\rm MW} & = &  \sum_{i=1}^{2} PL(E;C_i,\alpha_i) + \nonumber\\
    & + &\sum_{j=1}^{5} G(E;F_j,\mu_j,\sigma_j) + \nonumber\\
    & + & Ps(E;F_{\rm Ps},\mu_{\rm Ps},\sigma_{\rm Ps},f_{\rm Ps})\mrm{,}
    \label{eq:total_spectral_model}
\end{eqnarray}
where
$PL(E;C,\alpha)$ is a power law with normalisation $C$ and index $\alpha$, $G(E;F,\mu,\sigma)$ is a Gaussian with flux $F$, centroid $\mu$, and width $\sigma$, and $Ps(E;F_{\rm Ps},\mu_{\rm Ps},\sigma_{\rm Ps},f_{\rm Ps})$ is the Positronium spectrum with a $\mu_{\rm Ps} = 511$\,keV line of width $\sigma_{\rm Ps} = 1.7$\,keV plus ortho-Positronium, normalised by the line flux $F_{\rm Ps}$, taking into account a fraction, $f_{\rm Ps}$, of positrons annihilating via the Positronium channel \citep{Ore1949_511}.
The two power laws are fitted independently, taking into account the resolved and unresolved point sources at lower energies, and the IC emission at higher energies.
The five Gaussian lines describe the nuclear decay $\gamma$-rays of radioactive isotopes in the Milky Way, that are \nuc{Al}{26}, \nuc{Fe}{60} ($\times 2$), \nuc{Na}{22}, and \nuc{Be}{7}, and are fixed at their respective lab energies of 1809, 1173 and 1332, 1275, and 478\,keV, respectively.
The line widths are chosen as instrumental resolution, except for the 478 and 1275\,keV lines from classical novae with a broadening of $2000\,\mrm{km\,s^{-1}}$.
The line fluxes of the \nuc{Al}{26} and \nuc{Fe}{60} lines are linked by the measured Galactic flux ratio of $F_{60}/F_{26} = 0.184 \pm 0.042$ \citep{Wang2020_Fe60}, and the nova lines from \nuc{Be}{7} and \nuc{Na}{22} are allowed to vary within their uncertainties, that is, their upper limits of $< 6 \times 10^{-4}\,\mathrm{ph\,cm^{-2}\,s^{-1}}$ and $< 4 \times 10^{-4}\,\mathrm{ph\,cm^{-2}\,s^{-1}}$, respectively ($3\sigma$) \citep{Siegert2021_BHMnovae}.
The free parameters of the Galactic diffuse emission without the DM  component in the spectral fits, ignoring those which are bound to their priors, are hence, $C_1$, $\alpha_1$, $C_2$, $\alpha_2$, $F_{\rm Ps}$, $f_{\rm Ps}$, and $F_{26}$.

\begin{figure}
    \centering
    \includegraphics[width=\columnwidth]{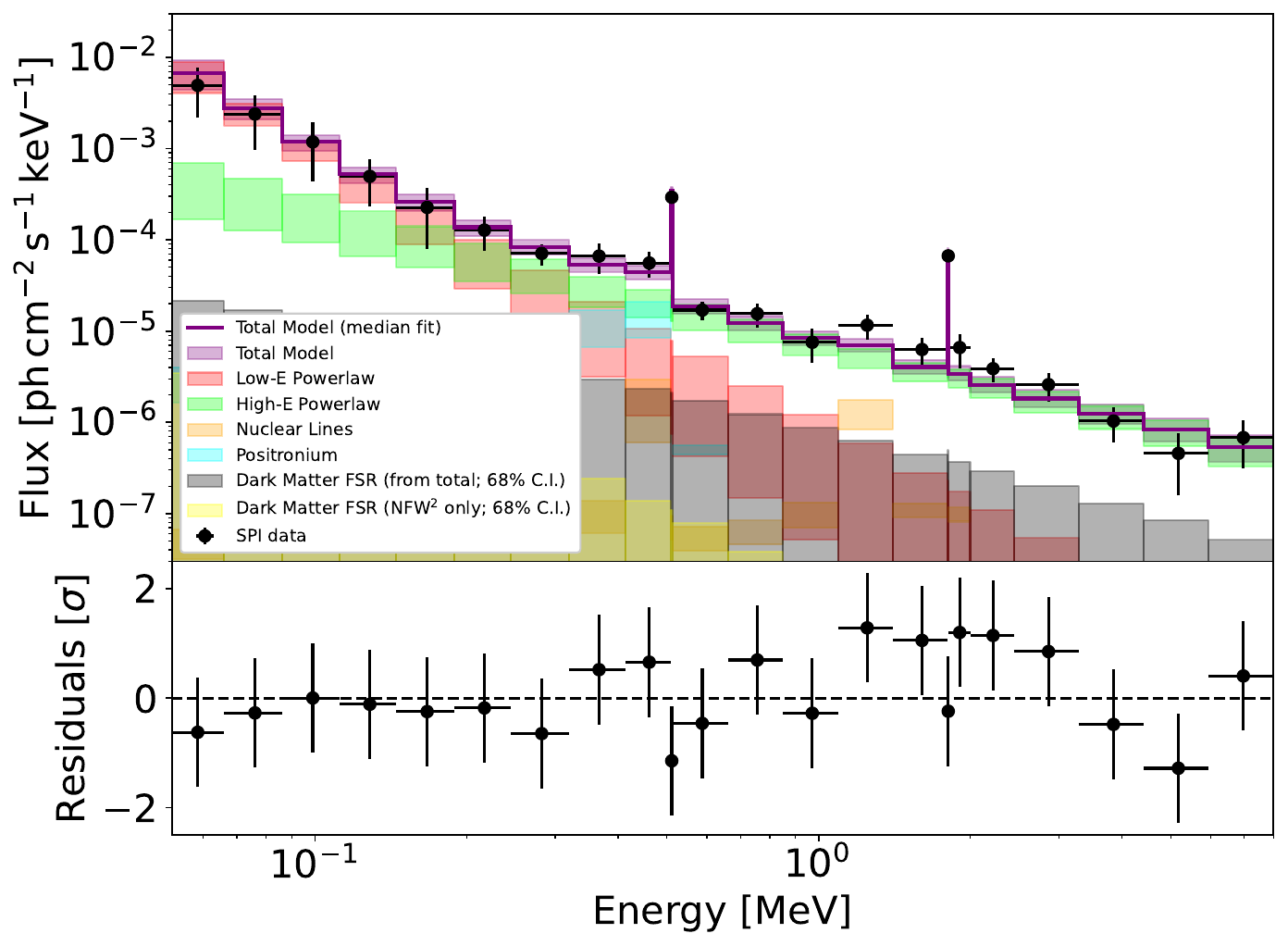}\\
    \includegraphics[width=\columnwidth]{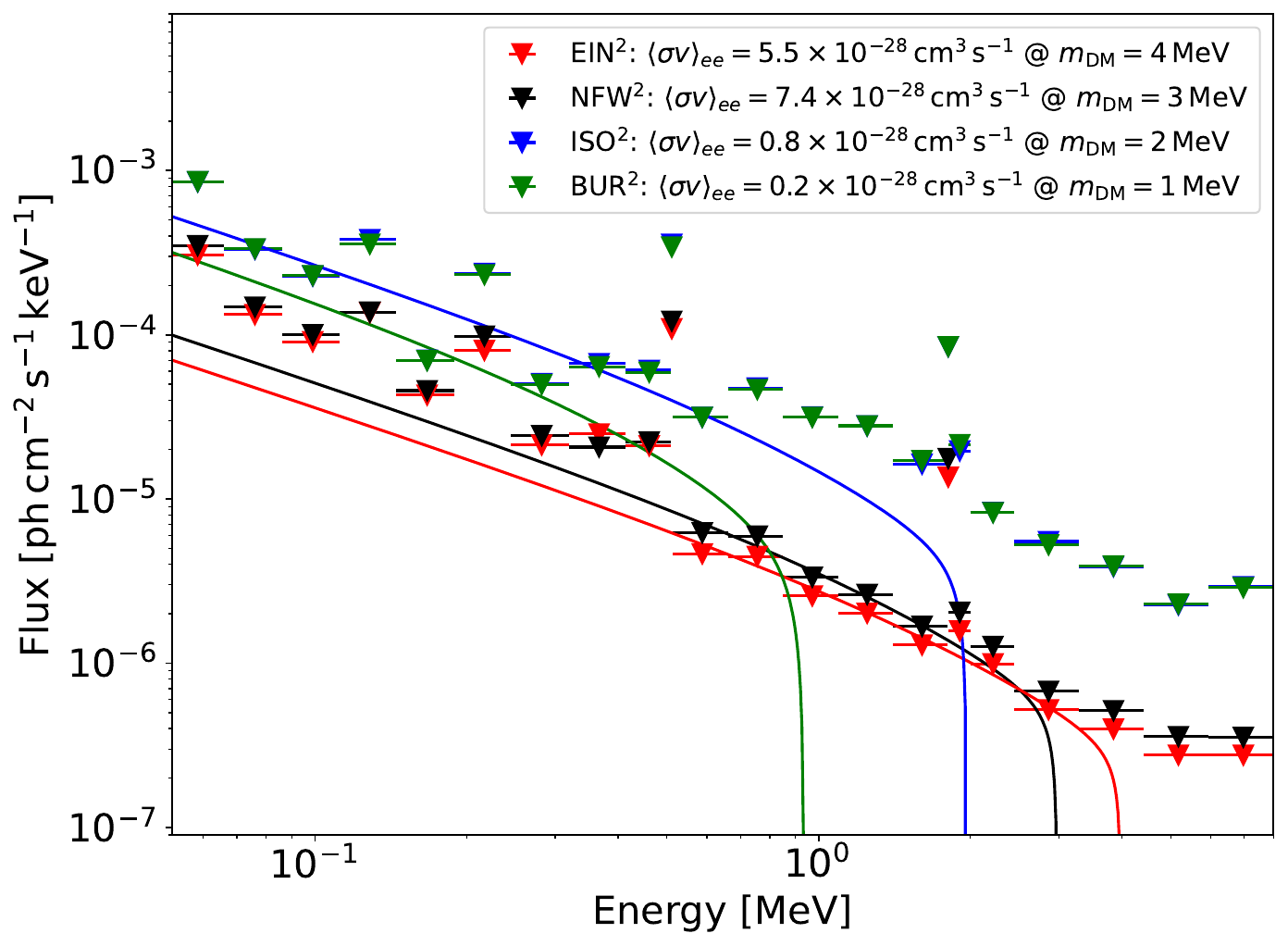}
    \caption{Top: Extracted ``total'' SPI spectrum from 16\,yr of INTEGRAL observations (black data points) including an NFW$^2$-profile, together with fitted models and fit residuals (sub panel). Bottom: Dark matter only spectra. Shown are $2\sigma$ upper limits for all used DM density profiles and some excluded FSR spectra in each case between DM masses of 1 and 4\,MeV, together with the excluded cross section (see legend).}
    \label{fig:total_spectrum}
\end{figure}

\subsubsection{Dark matter spectra}
The DM annihilation channels used in this work are either to a pure photonic final state, or the final state radiation (FSR) radiation from a pure leptonic (electron-positron) final state,
\begin{enumerate}
    \item $\mrm{DM + DM} \longrightarrow \gamma + \gamma$ (named $\gamma\gamma$), and
    \item $\mrm{DM + DM} \longrightarrow e^+ + e^- + \gamma_{\rm FSR}$ (named $ee$).
\end{enumerate}

The leptonic channel is suppressed by $\mathcal{O}(\alpha)$ with respect to tree-level annihilation processes for which the s-wave thermal relic cross section in the GeV mass range or above is $\approx 3 \times 10^{-26}\,\mrm{cm^{3}\,s^{-1}}$;
for lower masses such as considered here, this value is somewhat higher, with values up to $\approx 5 \times 10^{-26}\,\mrm{cm^{3}\,s^{-1}}$ for masses around $\approx 300$\,MeV \citep{Steigman2012_thermalrelic}.
Since DM is `dark', the photonic final state can only arise at loop level.
However, in the sub-MeV range, this channel may be the leading one in Standard-Model final states due to the lack of other kinematically available states.
The spectral shape, normalised by the self-annihilation cross section and $J$-factor of the Milky Way in the region of interest, also simply known as the DM spectrum, is given by
\begin{equation}\label{eq:flux_gg}
	\left(\frac{dN}{dE}\right)_{\gamma\gamma} = 2 \delta(E-m_{\rm DM})
\end{equation}
for case 1, and
\begin{equation}
\tiny{
    \left(\frac{dN}{dE}\right)_{ee\gamma} = \frac{\alpha}{2\pi}\left[\frac{(2m_{\rm DM})^2 + (2m_{\rm DM} - E)^2}{(2m_{\rm DM})^2E}\ln\left(\frac{2m_{\rm DM}(2m_{\rm DM}-E)}{m_e^2}\right)\right]}
    \label{eq:flux_FSR}
\end{equation}
for case 2 \citep{Beacom2005_DM_IB}.
We note that the expression used for case 2 only provides the dominant contribution in some limiting cases, such as the soft photon limit, as pointed out by \citet{2006hep.ph....6058B}.
We implicitly assume its validity for the sake of comparison with results in the literature, where this ``universal'' spectrum is typically adopted.
We further note that in case 2, the suppression factor due to the radiative nature of the channel is already factored in in the DM spectrum.

The total DM $\gamma$-ray flux for channel $c$ and self-conjugated particles thus reads
\begin{equation}
    \left(\frac{dN}{dE\,dA\,dt}\right)_{c} =  \frac{J \langle \sigma v\rangle}{4 \pi m_{\rm DM}^2} \left(\frac{dN}{dE}\right)_{c}\mrm{,}
    \label{eq:astro_spec}
\end{equation}
where $\langle \sigma v \rangle$ is the velocity-averaged self-annihilation cross section and $m_{\rm DM}$ is the DM particle mass.

\section{Analysis results}\label{sec:results}
\subsection{Spatial decomposition}
Adding a spatial template that resembles a DM annihilation halo to the astrophysically known components results in no significant detection ($2\sigma$) for any of the chosen profiles.
We show the total spectrum of the Milky Way between 0.05--8\,MeV including an NFW$^2$ profile in Fig.\,\ref{fig:total_spectrum} with systematic uncertainties derived from different IC emission models.
The spectral fit is shown in data space (top; taking into account the spectral response function of SPI) including the case of FSR from an NFW$^2$-profile as an example, with their $1\sigma$ uncertainty bands for clarity.
The spectra resulting from the DM$^2$-profiles alone is shown in the bottom panel of Fig.\,\ref{fig:total_spectrum}.
Since no flux has been detected for either of the DM profiles, we show the $2\sigma$ upper limits on the flux.

\subsection{Spectral fit and limits on dark matter}
\begin{figure}
    \centering
    \includegraphics[width=\columnwidth]{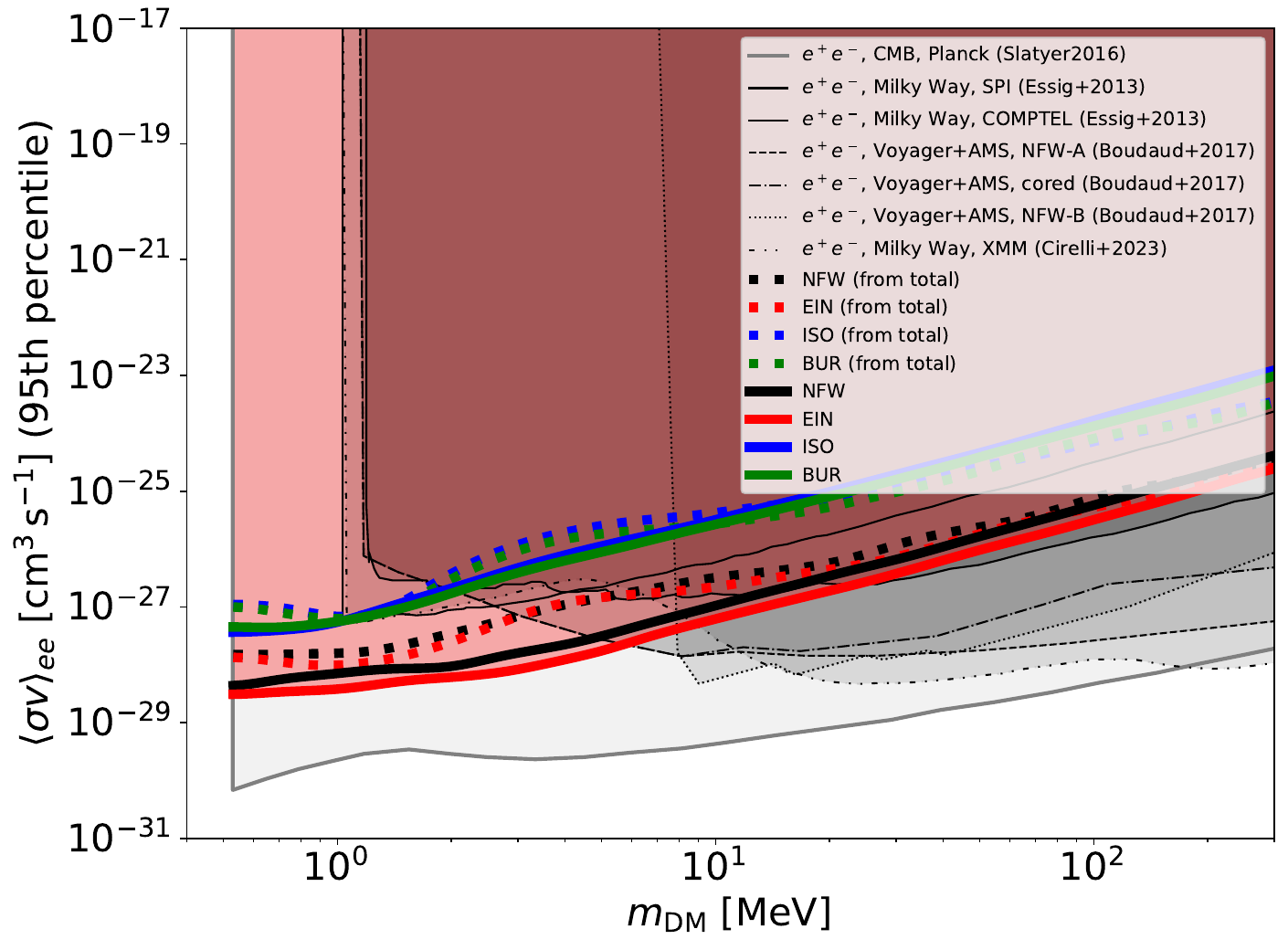}
    \caption{Upper bounds (95\% C.I.) on the velocity-averaged annihilation cross section of DM particles into the electron-positron final state, undergoing FSR as a function of particle mass. Either the total Galactic spectrum is used for spectral fits (`from total', dotted lines) or the respective dark-matter-only spectrum (solid lines). For comparison, the limits from CMB measurements are shown in gray \citep{Slatyer2016_CMB_DM}, together with limits from COMPTEL \citep{Essig2013_DMlimits_gammarays}, Voyager+AMS \citep{2017PhRvL.119b1103B}, and XMM-Newton \citep{2023JCAP...07..026C}.}
    \label{fig:FSRee_limits}
\end{figure}

We search for a residual spectral enhancement in the analysed energy band with two approaches:
a) Either, the total spectrum, that is, all spatial components including a DM halo and point sources, is fitted with the function in Eq.\,(\ref{eq:total_spectral_model}) plus Eq.\,(\ref{eq:astro_spec}) to find a possible DM component (`from total' in Figs.~\ref{fig:FSRee_limits} and \ref{fig:gg_limits_from_total}; see spectrum in Fig.\,\ref{fig:total_spectrum}), or b) the DM-only spectrum (mostly consistent with zero within $2\sigma$ in all energy bins) is fitted with Eq.\,(\ref{eq:astro_spec}) alone.

\begin{figure}
    \centering
    \includegraphics[width=\columnwidth]{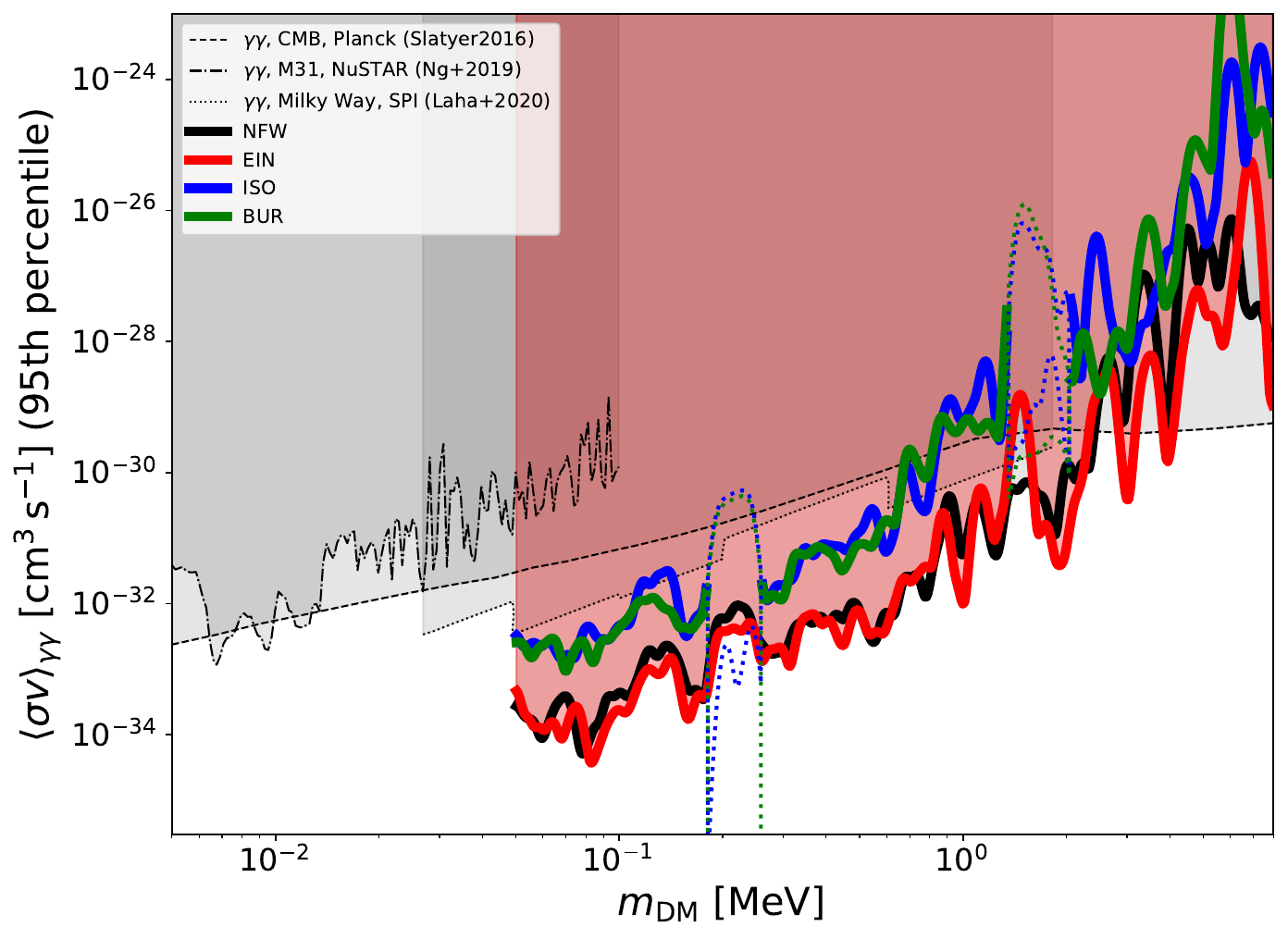}
    \caption{Same as Fig.\,\ref{fig:FSRee_limits} but for the two-photon annihilation channel. For better readability, the $\gamma$-ray line search from the total spectrum is shown in Appendix, Fig.\,\ref{fig:gg_limits_from_total}. ISO and BUR limits show $2\sigma$ indications at the positions of strong instrumental background lines (thin dotted coloured lines).}
    \label{fig:gg_limits}
\end{figure}

For each of the four different DM halo profiles, we perform a separate analysis for the $ee$ and $\gamma\gamma$ final states.
We detect no sign of an additional FSR or $\gamma$-ray line emission that would originate from a DM halo.
While the FSR spectrum is hardly compatible with the measured data, the narrow-line search results in a few $2\sigma$ excesses, but only in the case of an ISO or BUR profile.
The photon energies at which these excesses are seen are around 75, 198, and 1779\,keV, reminiscent of instrumental background lines in the SPI raw data \citep{Diehl2018_BGRDB}.
We discuss the correlations of celestial emission and background emission in more detail in Sec.\,\ref{sec:discussion}.

We provide upper bounds on the velocity-averaged annihilation cross section $\langle \sigma v \rangle$ for both annihilation channels for all DM profiles discussed in Sec.~\ref{sec:halos}.
The bounds are shown as a function of DM mass in Fig.\,\ref{fig:FSRee_limits} for the electron-positron channel, and in Fig.\,\ref{fig:gg_limits} for the two-photon channel.
Since the behaviour of the bounds are rather smooth with expected variations from the narrow-line model (Fig.\,\ref{fig:gg_limits}), we can describe our bounds with a simple scaling equation.
We use the bounds derived from the NFW profile as a benchmark here.
Bounds for other DM halo profiles behave similarly in the $ee$ channel but add some structure in the $\gamma\gamma$ channel (see Sec.\,\ref{sec:discussion} for discussions).
If we describe the 95\% C.I.~upper bound on $\langle \sigma v \rangle$ with a broken power law, we find that $\langle \sigma v \rangle_{ee}$ scales as $\leq 0.7 \times 10^{-28}\,\left(\frac{m_{\rm DM}}{\mrm{MeV}}\right)^{\epsilon}\,\mathrm{cm^{3}\,s^{-1}}$ with $\epsilon = 0.6$ for $m_{\rm DM} \leq 3$\,MeV and $\epsilon = 1.7$ up to 300\,MeV (maximum of tested range).
For the two-photon annihilation channel, $\langle \sigma v \rangle_{\gamma\gamma}$ scales as $\leq 3 \times 10^{-33}\,\left(\frac{m_{\rm DM}}{\mrm{MeV}}\right)^{\nu}\,\mathrm{cm^{3}\,s^{-1}}$ with $\nu = 1.1$ for $m_{\rm DM} \leq 0.25$\,MeV and $\nu = 4.0$ up to $m_{\rm DM} = 8$\,MeV.
In this case, our analysis finds the strongest bounds to date in the range from 50\,keV to $\sim 1.5$\,MeV, almost independently on the DM halo profile, and superseding CMB limits \citep{Slatyer2016_CMB_DM} by almost two orders of magnitude.
In the corresponding figures, we compare our results with current limits from the literature, in particular from \citet{Slatyer2016_CMB_DM} using CMB measurements, \citet{Ng2019_DM_NuSTAR_M31} using NuSTAR measurements of M31,  \citet{Laha2020_PMBHDM} using already-extracted INTEGRAL/SPI data, and \citet{2017PhRvL.119b1103B} using Voyager I and AMS-02 cosmic-ray data.
Other competitive limits for higher masses come from the inclusion of IC emission in leptonic DM annihilation channels \citep{2021PhRvD.103f3022C,2023JCAP...07..026C}.
For illustration purpose, we report the strongest limits available in the literature.

\begin{figure*}
    \includegraphics[width=0.32\textwidth]{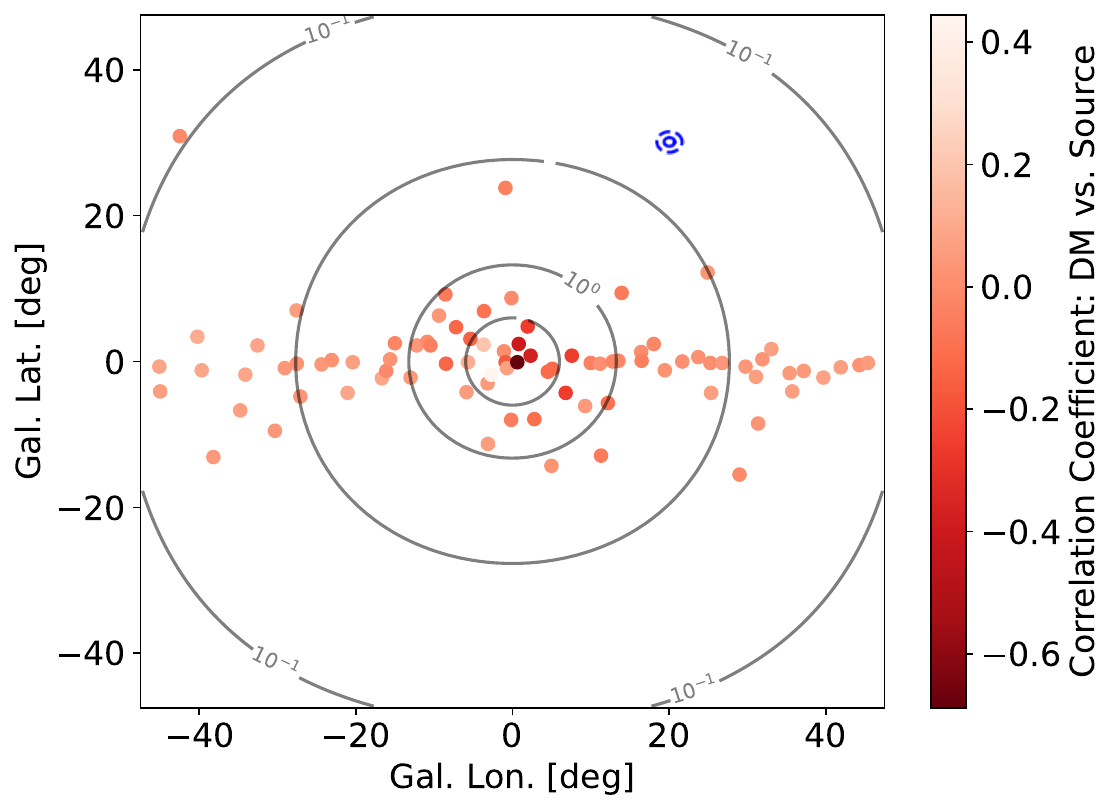}
    \includegraphics[width=0.32\textwidth]{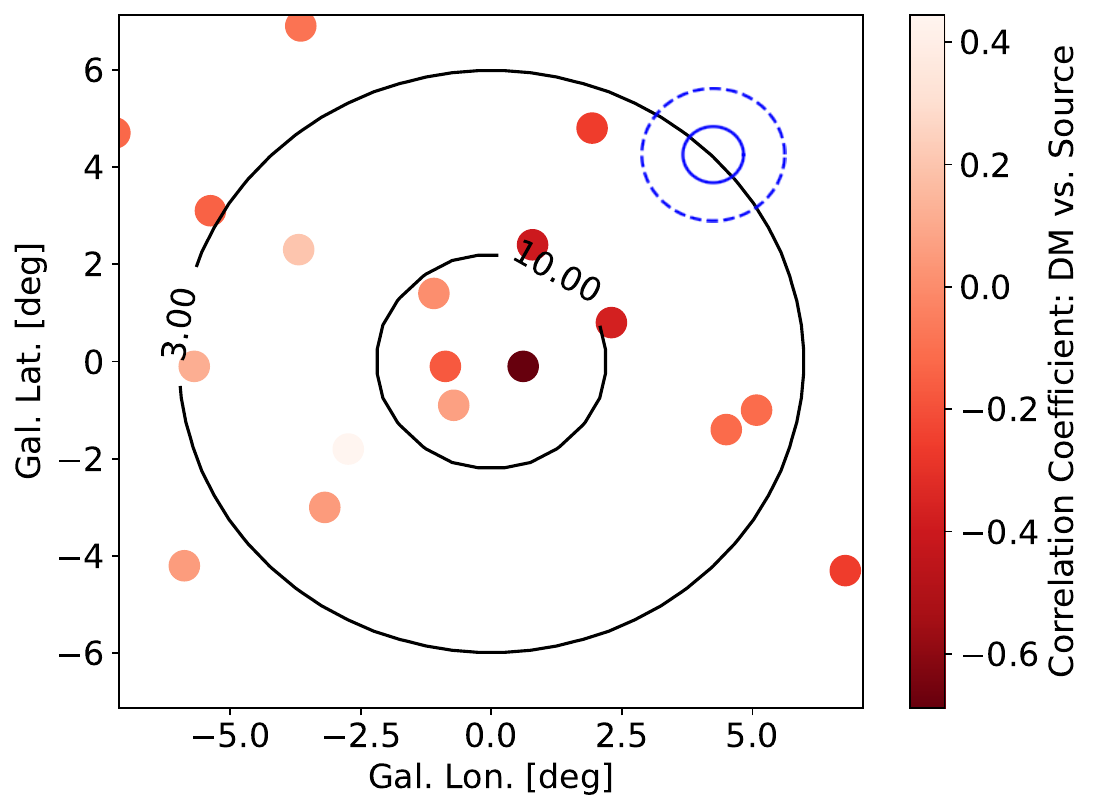}
    \includegraphics[width=0.32\textwidth]{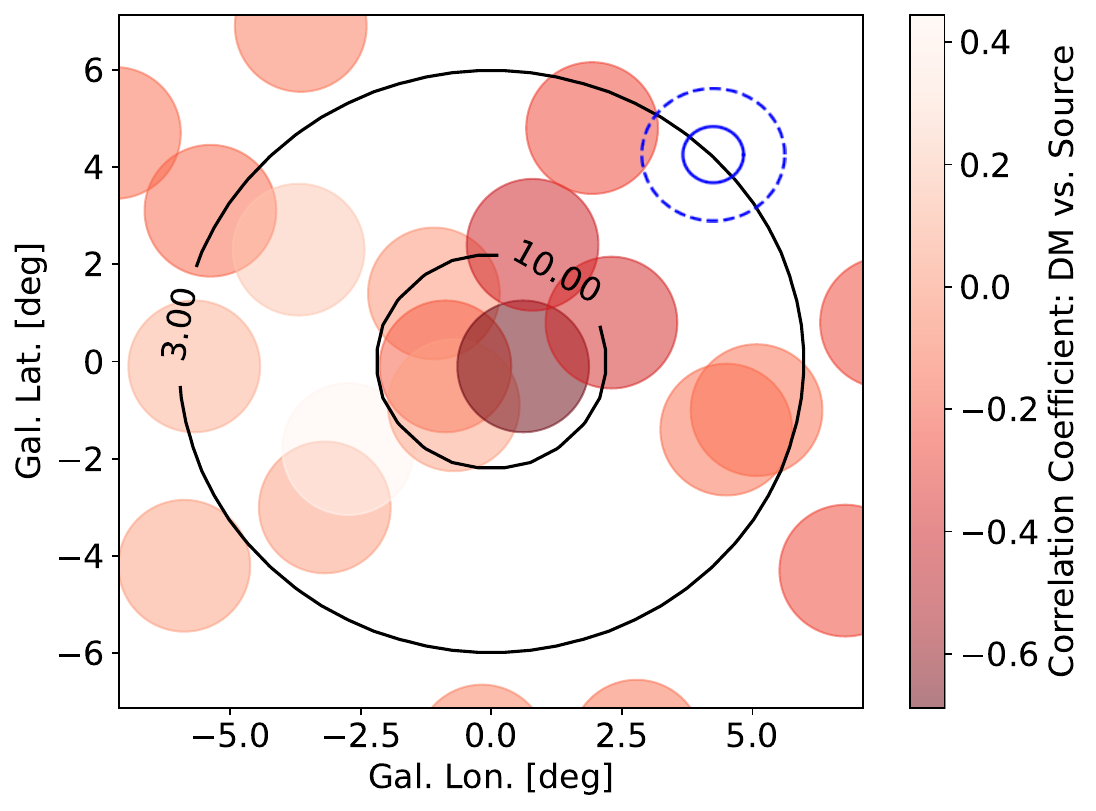}
    \caption{Spatial correlation of an NFW$^2$-profile (contours) with 92 significant point sources in the energy band 51--66\,keV. From left to right: All sources in the region of interest, zoom in to a region with the strongest correlations ($|\ell| \leq 7.5^\circ$, $|b| \leq 7.5^\circ$), and the same zoom in but with point source sizes of one SPI PSF (marked as blue circles at 1$\sigma$ [solid line], and FWHM [dashed line]).}
    \label{fig:NFW_correlations_spatial}
\end{figure*}

\section{Discussion and conclusions}\label{sec:discussion}
\subsection{Spatial correlations}\label{sec:spatial_correlations}
The data analysis of a coded-mask telescope, such as INTEGRAL/SPI, relies on the coding pattern received from different sources with respect to the observation pattern, that is, it requires contrast between different positions in the sky.
The more sources (point-like and diffuse) contribute in a given energy bin, the higher is the fitted covariance among these emission templates, resulting in source confusion.
This means, the extracted fluxes for individual components can be under- or over-estimated, whereas the total emission is still bound to the total number of photons recorded.
This problem grows particularly strong when the diffuse emission component is highly peaked and overlapping with many point sources, such as in the case of the NFW profile (squared) in the Galactic centre.
We show the spatial correlation of 92 point sources in the energy bin 51--66\,keV with the NFW$^2$-profile in Fig.\,\ref{fig:NFW_correlations_spatial} as a function of Galactic longitude and latitude.
It is evident that in the wings of the DM halo profile, the sources are not affected (Fig.\,\ref{fig:NFW_correlations_spatial}, left), however the (anti)-correlations become considerably large in the central $|\ell| \leq 7.5^{\circ}$ and $|b| \leq 7.5^{\circ}$ (Fig.\,\ref{fig:NFW_correlations_spatial}, middle).
This is due to the fact that the angular resolution of SPI is about $2.7^\circ$, which leads to large overlaps between the many sources individually, as well as the NFW$^2$-profile as a whole.
To alleviate the problem, we derive the spectrum of the Milky Way Galaxy as a whole when including a DM halo profile squared, from which we can derive bounds on DM particle properties without large biases.
These bounds are then about a factor of $\sim 2$ weaker in the $ee$ channel (or stronger for higher DM masses, see Fig.\,\ref{fig:FSRee_limits}), and up to a factor of $\sim 10$ weaker in the $\gamma\gamma$ channel.
Still, the limits on $\langle \sigma v \rangle_{\gamma\gamma}$ are the strongest in the literature between 50\,keV and $\sim 1.5$\,MeV.

The $2\sigma$ excesses visible in Fig.\,\ref{fig:gg_limits} are seen at energies around 75, 198, and 1779\,keV, for only the ISO and BUR profiles.
In these cases, the instrumental background lines of SPI `shine through' and the fitting algorithm \citep[based on a maximum likelihood method,][]{Siegert2019_SPIBG,Siegert2022_MWdiffuse} places background into celestial components which mimic background.
This can be understood as follows:
As mentioned above, coded-mask telescopes require contrast between different components to distinguish them.
The ISO and BUR profiles only show very weak gradients (Fig.\,\ref{fig:2D_profiles}, two right figures), even when the profiles are squared.
This leads to the effect that with the observational strategy of SPI, dithering only $2.1^\circ$ from one pointed observation to the next \citep{Vedrenne2003_SPI}, the BUR and ISO imprints on the detector array through the mask coding are almost identical for every direction.
An identical detector imprint, however, is what describes the instrumental background, so that here, the correlation between background and a shallow-gradient emission template is large, and the flux might be wrongly attributed.
Since this only occurs for the case of the ISO and BUR profiles and not for the EIN and NFW profiles, we can be certain that the $2\sigma$ line excesses are due to instrumental background.

\subsection{Comparison to literature}\label{sec:literature_comparison}
Limits on $\langle \sigma v \rangle_{ee}$ are available from CMB measurements \citep{Slatyer2016_CMB_DM}, setting the most stringent bounds for thermal relic production, as well as from previous $\gamma$-ray measurements with SPI and COMPTEL \citep{Essig2013_DMlimits_gammarays}, and Voyager I electrons and positrons local measurements \citep{2017PhRvL.119b1103B}.
The work by \citet{Essig2013_DMlimits_gammarays} is based on already-extracted flux values so that the bounds may be largely over-optimistic.
In fact it was shown already in our previous works that deriving limits from existing data without taking into account the spatial distribution of the signal can lead to erroneous results, either over- or under-estimating the actual limits~\citep{Calore2023_lightDM}.
Voyager I limits \citep{2017PhRvL.119b1103B} can be quite competitive for masses above 10 MeV.
The same is true for limits which include IC emission in leptonic DM annihilation, and exploit X-ray data from XMM-Newton \citep{2023JCAP...07..026C}.
Compared with CMB limits which considered s-wave annihilation into electron-positron pairs, our bounds are one to two orders of magnitude weaker for the NFW or EIN profiles, and two to three orders of magnitude weaker for the BUR or ISO profiles.

Recent limits on the velocity-averaged cross section of DM particles annihilating into two photons, $\langle \sigma v \rangle_{\gamma\gamma}$, have been derived by \citet{Ng2019_DM_NuSTAR_M31} from NuSTAR observations of M31 in the mass range up to $\sim 100$\,keV, by \citet{Laha2020_PMBHDM} using already-extracted SPI data between 0.03 and 1.8\,MeV, and again the CMB limits from \citet{Slatyer2016_CMB_DM}. 
While the previous SPI limits only consider data in which no DM assumption is embedded \citep{Laha2020_PMBHDM}, our method includes all our previous knowledge about the astrophysical background emission processes, the morphology of the DM signal, as well as a measure of systematic uncertainties.
Our limits are one to two orders of magnitude better than the previous SPI limits and two to three orders of magnitude better than the CMB limits, depending on the used DM halo profile.
We therefore strongly constrain thermal relic particles in the mass range between 0.05 and 3\,MeV, extending the previous findings.
We further show that, even though the BUR and ISO profiles are difficult to handle within the raw data analysis framework of INTEGRAL/SPI (see Sec.\,\ref{sec:spatial_correlations}), they are also excluded to accommodate light DM particles in the Milky Way annihilating into two photons up to a mass range of $\sim 1$\,MeV.
These exclusions are robust because both our analysis approaches, either fitting the total spectrum or fitting the DM-only spectrum, result in upper bounds on the two-photon annihilation channel much below $10^{-30}\,\mathrm{cm^{3}\,s^{-1}}$ for DM particle masses between 0.05 and 1\,MeV, that is, the benchmark tree-level cross section for thermal relics proceeding via $\mrm{DM + DM} \longrightarrow e^+ + e^-$, if kinematically open.

We note that, if DM particles annihilate through velocity dependent cross section for instance, p-wave limits from the Galactic diffuse emission cannot, at present, exclude thermal production, $\langle \sigma v \rangle_{\rm th, p} \sim 10^{-32}\, \mathrm{cm^{3}\,s^{-1}}$, in any of the two channels tested here.
This is even more true for CMB limits, which vanish almost completely for p-wave processes.
Assuming the velocity dispersion of the Milky Way to be of the order of $v_r \sim 300\,\mathrm{km\,s^{-1}}$, the velocity-dependent part of the annihilation cross section will obtain an additional term that roughly follows \citep{2019PhRvD..99f1302B}
\begin{eqnarray}
    & & \langle \sigma v \rangle = \nonumber\\
    & & \langle \sigma v \rangle_{\rm s-wave} + \langle \sigma v \rangle_{\rm p-wave} + \mathrm{higher\,order\,terms} = \nonumber\\
    & & \sigma_0 c + \sigma_1 c \left(\frac{v_r}{c}\right)^2 + \dots\mrm{.}
    \label{eq:cross_section_expansion}
\end{eqnarray}
The p-wave limits therefore scale as $10^{-6}$ of those of the s-wave limits.
Considering the estimates from \citet{2019PhRvD..99f1302B}, our p-wave limits would fall in the same order of magnitude as those, between $\sigma_1 c \lesssim 10^{-23}$--$10^{-21}\,\mathrm{cm^3\,s^{-1}}$ in the mass range from a few MeV to a few GeV.
Hence, while far from theoretically interesting targets, we expect INTEGRAL/SPI to still provide competitive if not the best bounds below $\sim10$\, MeV.
Given the scarce interest for this parameter space, we leave the proper consideration of realistic velocity distributions of the DM particles in the Milky Way halo for future work.

One could improve the limits especially in the low-mass range, below $\sim 4$\,MeV if the diffuse IC scattering emission in the Milky Way was better understood, so that its spectral shape could be better constrained in the analysis.
For the tree-level annihilation into $ee$, this can lead to limits below 1\,MeV of $\lesssim 10^{-29}\,\mathrm{cm^{3}\,s^{-1}}$ even with SPI.
Using the entire sky and the now more than 20 years of observations may even improve the bounds further.
However, with current instrumentation, it may be impossible to beat the CMB limits with Milky Way observations of INTEGRAL/SPI alone.
The future COSI-SMEX satellite mission, planned for launch in 2027, will enhance the sensitivity in the MeV range thanks to its large field of view and improved performance \citep{Tomsick2019_COSI,Caputo2023_COSI_DM,2023arXiv230812362T}.

\section*{Acknowledgements}
This work is supported by the ``Agence Nationale de la Recherche”, grant n. ANR-19-CE31-0005-01 (PI: F. Calore).
We thank Saurabh Mittal for thoroughly reading the manuscript.

\section*{Data Availability}
Raw and extracted data will be made available upon reasonable request.



\bibliographystyle{mnras}
\bibliography{thomas} 

\appendix

\section{Correlations between components}\label{sec:appendix_correlations}
As discussed in Sect.\,\ref{sec:spatial_correlations}, there are correlations between the different sky (and background) components due to the spatial overlap of sources.
For a better understanding of how strong the correlations can actually be, we show the correlation matrix for two different energy bins when using the NFW$^2$-- or the BUR$^2$--profile in Appendix Fig.\,\ref{fig:correlation_matrix}.
In the top figure, in the energy band from 51 to 66\,keV, there are four extended source components (indices 0 to 3) and 92 point-like sources (indices 4 to 95).
The point source index is indicating the longitudinal position ranging from $\ell = -47.5^\circ$ to $\ell = +47.5^\circ$ in the indicated index range.
It is evident that point sources that are close to the Galactic centre and bulge have stronger correlations with each other and with the diffuse emission templates than sources further out in the disk.
Clearly, the deviations along the diagonal are correlations of sources in the vicinity to each other.
Likewise, in the bottom figure, in the energy band from 189 to 245\,keV, there are only two extended emission components and 21 point sources.
The general structure is the same and anti-correlations with the Positronium component (index 1) are clearly dominating in the Galactic bulge region (indices 9--17).

\begin{figure}
    \centering
    \includegraphics[width=\columnwidth]{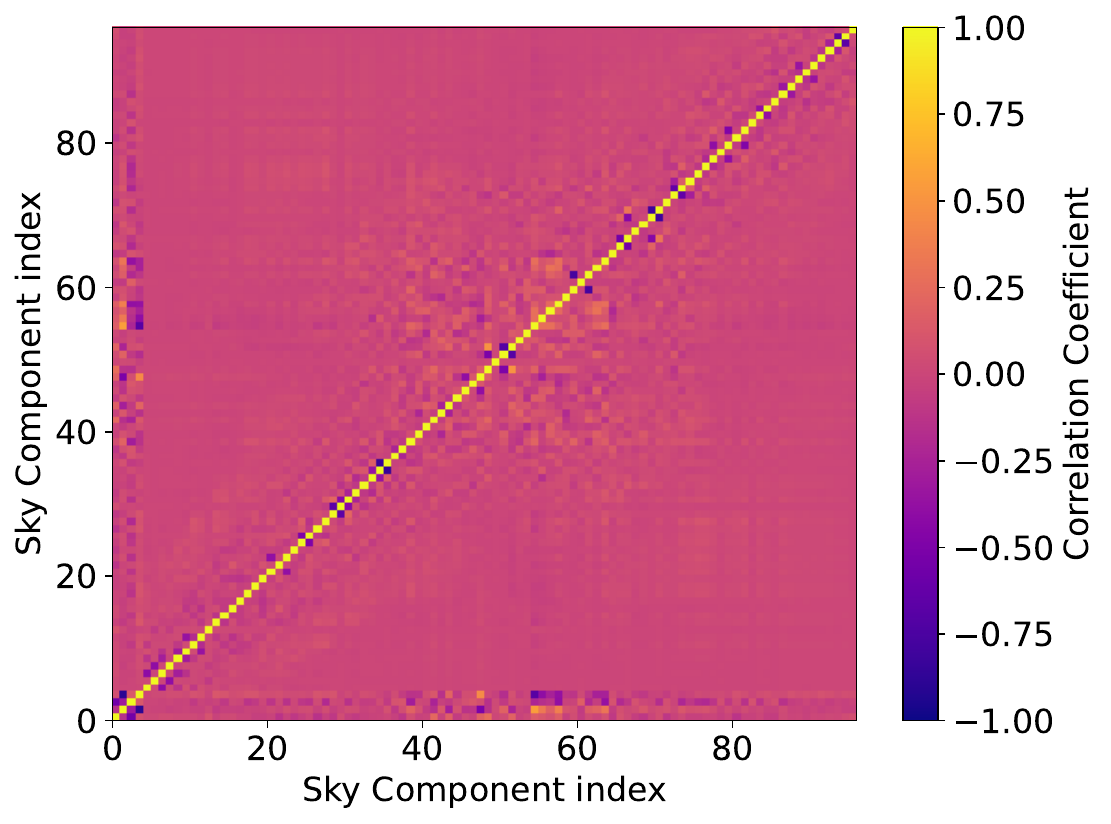}\\
    \includegraphics[width=\columnwidth]{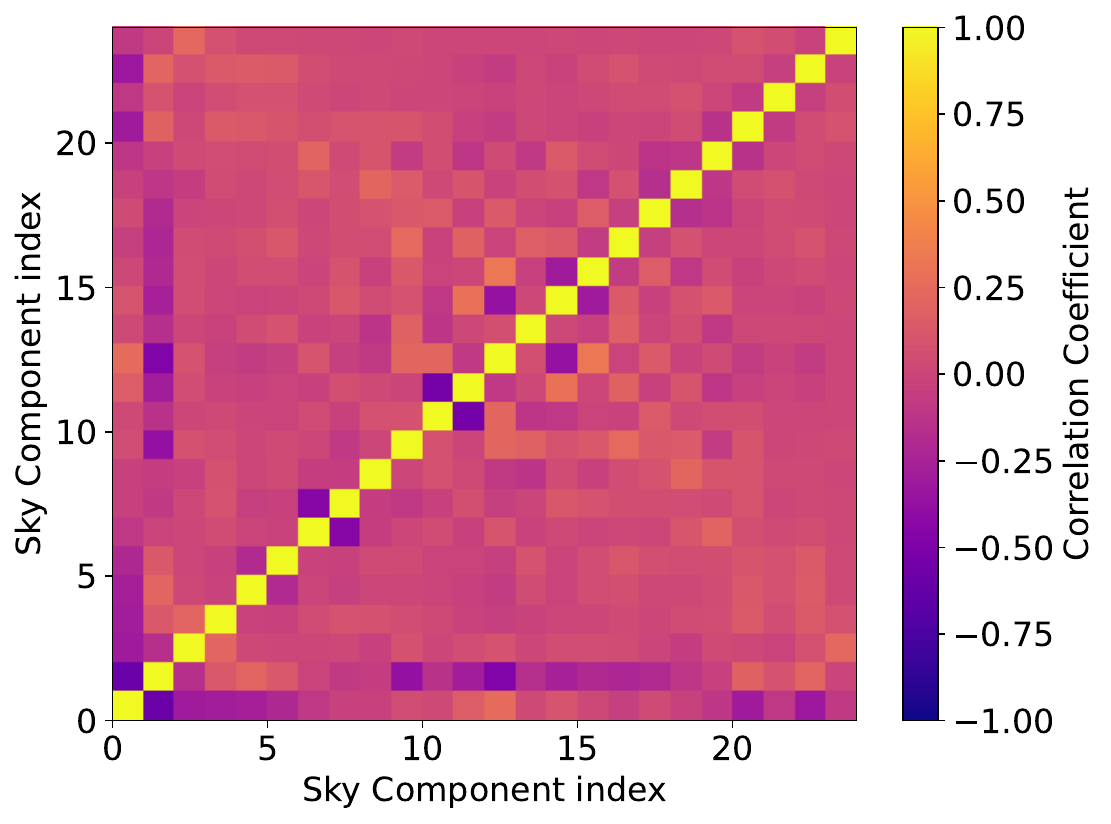}
    \caption{Correlation matrix between sky model components. \textit{Top:} The first four entries correspond to the extended emission models of cataclysmic variables, Positronium emission, IC emission ($\delta_1 = \delta_2 = 0.5$), and an NFW$^2$-profile, respectively, in the energy band between 51 and 66\,keV, whereas the remaining components are 92 point-sources (cf. Fig.\,\ref{fig:NFW_correlations_spatial}). \textit{Bottom:} Same as above, but for the three extended components of Positronium, IC (Voyager baseline), and a BUR$^2$-profile, respectively, in the energy band between 189 and 245\,keV with 21 sources.}
    \label{fig:correlation_matrix}
\end{figure}


\section{Additional plots}\label{sec:additional_plots}
\begin{figure}
    \centering
    \includegraphics[width=\columnwidth]{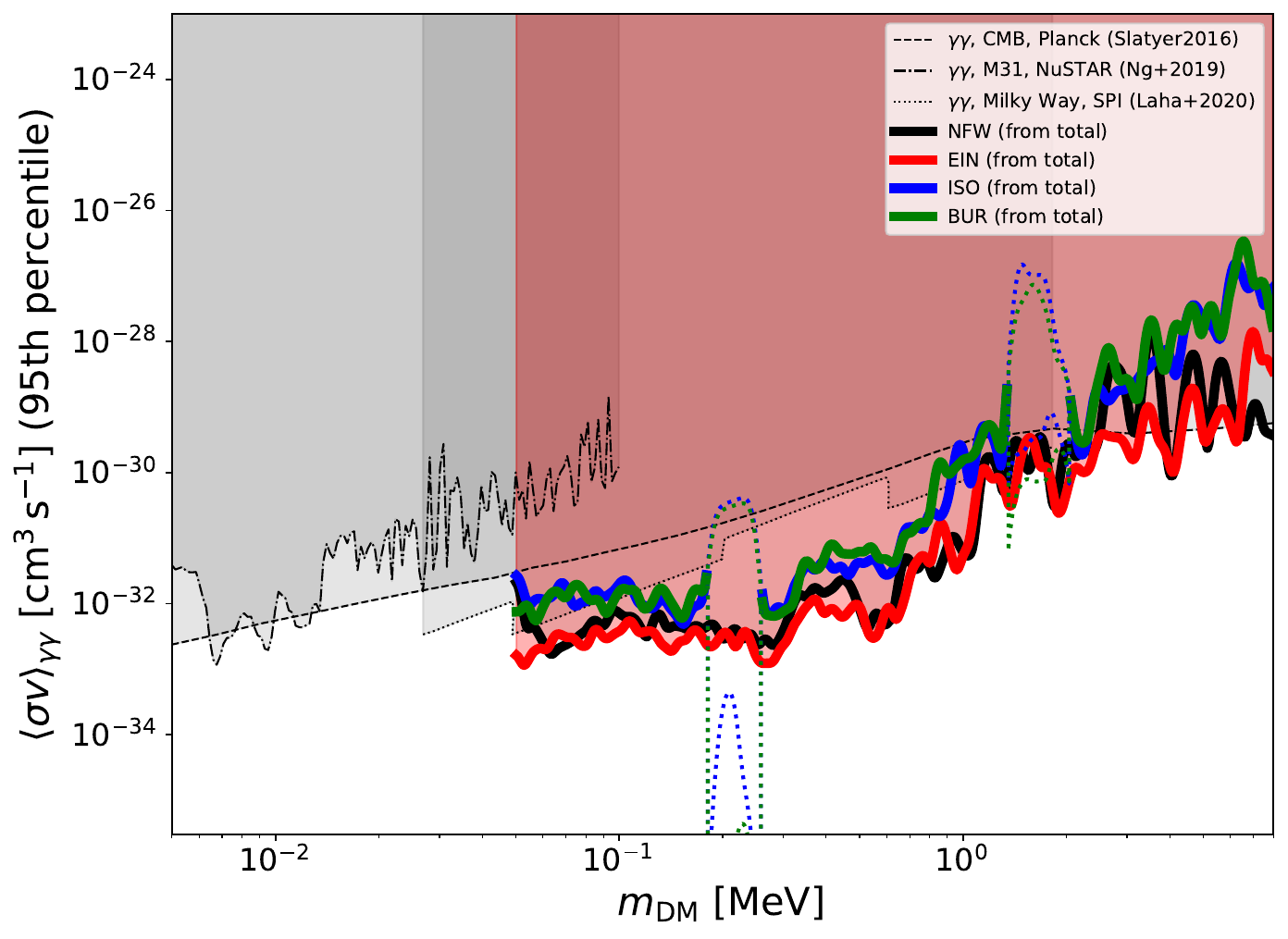}
    \caption{Same as Fig.\,\ref{fig:gg_limits} but with limits derived from the total Galactic emission spectrum.
    }
    \label{fig:gg_limits_from_total}
\end{figure}

For completeness, we show the upper bounds on the velocity-averaged cross section into two photons derived from fits of the total Milky Way spectrum in Fig.\,\ref{fig:gg_limits_from_total}.

\bsp	
\label{lastpage}
\end{document}